\documentclass[nofootinbib,aps,letterpaper,superscriptaddress,showpacs,preprintnumbers]{revtex4}
\usepackage{amsmath}
\usepackage{dsfont}
\usepackage[dvips]{graphicx}

\usepackage{amsfonts}
\usepackage{amssymb}
\usepackage[dvips]{color}

\begin{document}
\preprint{IHES/P/09/28}
\newcommand\be{\begin{equation}}
\newcommand\ee{\end{equation}}
\newcommand\bea{\begin{eqnarray}}
\newcommand\eea{\end{eqnarray}}
\newcommand\bseq{\begin{subequations}} 
\newcommand\eseq{\end{subequations}}
\newcommand\bcas{\begin{cases}}
\newcommand\ecas{\end{cases}}
\newcommand{\p}{\partial}
\newcommand{\f}{\frac}

\title{On the gravitational polarizability of black holes}

\author{Thibault Damour}
\email{damour@ihes.fr}
\affiliation{Institut des Hautes Etudes Scientifiques, 35, route de Chartres, 91440 Bures-sur-Yvette, France.}
\author{Orchidea Maria Lecian}
\email{lecian@ihes.fr}
\affiliation{Institut des Hautes Etudes Scientifiques, 35, route de Chartres, 91440 Bures-sur-Yvette, France.}
\affiliation{APC, UMR 7164 du CNRS, Universit\'e Paris 7,\\10, rue A. Domon et L. Duquet, 75205 Paris Cedex 13, France.}
\affiliation{Sapienza- Universit\`a di Roma, Dipartimento di Fisica and ICRA, P.le A. Moro 5, 00185 Rome, Italy.}


\begin{abstract}
The \textit{gravitational polarizability} properties of black holes are compared and contrasted with their \textit{electromagnetic polarizability} properties. The ``shape'' or ``height'' multipolar Love numbers $h_l$ of a black hole are defined and computed. They are then compared to their electromagnetic analogs $h_l^{\rm EM}$. The Love numbers $h_l$ give the height of the $l$-th multipolar ``tidal bulge''  raised on the horizon of a black hole by faraway masses. We also discuss the shape of the tidal bulge raised by a test mass $m$, in the limit where $m$ gets very close to the horizon.
\end{abstract}


\maketitle

\section{Introduction}
In Newtonian gravity, the quantitative theory of the ``gravitational polarizability'' of elastic, self-gravitating bodies was pioneered by Love \cite{love1911}, who introduced two dimensionless measures of the response of an elastic body to an external tidal solicitation. To define them, let us first decompose the external tidal potential into multipolar components, say $U_{\rm ext}=\sum U_l=\sum T_l r^l P_l(\cos\theta)$, where the coefficient $T_l$ measures the strength of the $l$-th multipolar component of the external tidal field. The first ``Love number'', $h_l$, measures, essentially, the ratio between the $l$-th multipolar component of the distortion of the shape of the considered elastic body and $T_l$, while the second ``Love number'', $k_l$, measures the ratio between the $l$-th order multipole moment induced in the elastic body and $T_l$.\\
\\
In the membrane approach to black holes (BH's) \cite{Damour:1978cg,Znajek:1978,Damour:1979,Damour:1982,Macdonald:1985,Thorne:1986iy}, BH's are treated as elastic objects, endowed with usual physical properties. This raises the issue of defining and determining the BH analog of the Love numbers $h_l$ and $k_l$. Some time ago, Suen \cite{suen1986} made an attempt at defining and computing $k_2$, i.e. the quadrupole moment induced in a BH by an external quadrupolar tidal field. He surprisingly  found that, with his definition of the multipole moments of a distorted BH, the tidally induced quadrupole moment was \textit{opposite} to the externally applied quadrupolar tidal field (which would mean $k_2<0$, in contrast with usual elastic bodies, which have $k_l>0$). This unexpected result might well be due to the inappropriateness of the definition of induced multipole moments adopted in \cite{suen1986}. In two recent works, \cite{Damour:2009vw,Binnington:2009bb}, it was found that if one defines the $k_l$ Love number of a BH as being the mathematical continuation, up to a star compactness formally equal to $1/2$, of the (linearized) $k_l$ Love number of a neutron star, the resulting $k_l$ vanishes for all $l$'s. However, there are subtleties inherent in any definition of the multipole moments of BH's, so that there is currently no unambiguous determination of the $k_l$ Love number of BH's (see, e.g., \cite{Fang:2005qq} and \cite{Damour:2009vw} for discussions of the various ambiguities in the definition of the multipole moments of BH's). In the present paper, we shall not try to address these subtleties, and we shall, instead, focus on the computation of $h_l$, i.e. on the quantitative measure of the tidal distortion of the shape of a BH. By contrast to $k_l$, we shall see that there is no ambiguity in the computation of the first Love number $h_l$ of a BH.\\
Pioneering investigations of the tidal distortion of BH's were performed by Manasse \cite{manasse2} and Hartle \cite{hart1973,Hartle:1974gy} (see also \cite{Hawking:1972hy}). Further investigations are due to D'Eath, \cite{D'Eath:1975qs,D'Eath:1996nf}, and, more recently, to Poisson and collaborators \cite{Poisson:2005pi,Taylor:2008xy}. Though these papers explicitly discussed the quadrupolar ($l=2$) tidal distortion of BH's, they did  not consider higher multipolar orders of tidal distortion, nor did they explicitly compute the value of the $h_2$ Love number. In addition, there are sign errors in some formulas of \cite{hart1973}, which would affect the computation of $h_2$.\\
\\
The first purpose of this paper is to compute (in the linear approximation to tidal effects) the ``shape distortion'' Love numbers $h_l$ of a BH, for all values of $l\geq2$. Let us mention that a recent investigation of the Love numbers of neutron stars \cite{Damour:2009vw} has found that in the formal limit where the compactness $c^{NS}=GM/R$ of a neutron star tended to the compactness of a BH, $c^{BH}=1/2$, the $c$-dependent $h_l$ Love numbers of neutron stars tended to the BH $h_l$ values determined in this paper. The sequence of $h_l$'s is a way of parameterizing the tidal effects due to a disturbing mass $m$ located \textit{far away} from the considered BH. We shall also be interested in studying the distortion of the shape of a BH when a disturbing \textit{test} mass gets \textit{very near} the surface of the BH. In this respect, we will find it useful to compare and contrast the gravitational polarizability of BH's to their electric polarizability properties. Let us recall that Hanni and Ruffini \cite{Hanni:1973fn} pioneered the study of the electric polarizability properties of BH's, and introduced the notion of the charge density, $\sigma$, induced on the BH horizon by an external charge $q$. [This concept of induced charge density was one of the origins of the membrane approach to BH physics \cite{Damour:1978cg,Znajek:1978,Damour:1979,Damour:1982,Macdonald:1985,Thorne:1986iy}.] Below, we shall use the concept of induced charge density to define electric analogs of the $h_l$ Love numbers, which we shall compare to the gravitational ones. Then, we shall also compare the evolution of the gravitational, or electric, induced effects as the external tidally-influencing (test) mass $m$, or charge $q$, approach the horizon. [Ref \cite{Suen:1988kq} has considered the tidal distortion of the horizon of a BH by nearby (moving) masses, but has mainly used a ``Rindler approximation'' which replaces the BH horizon by a null hyperplane in a Minkowski background. In Section V, we shall discuss the connection of our results to those of \cite{Suen:1988kq}.]\\
\\
The paper is organized as follows. In Section II, the notion of gravitational shape Love number of a BH is introduced. It is defined as the dimensionless ratio between the $l$-th multipolar component of the deformation of the horizon geometry of a BH and the $l$-th multipolar component of the external tidal potential generated by a static axisymmetric distribution of faraway masses. In Section III, one introduces the notion of electromagnetic Love number by considering the influence of a distribution of faraway charges on the charge density induced on the surface of a BH. The case of a test charge approaching the BH horizon is examined in Section IV, while the case of an infinitesimal mass approaching the BH horizon is considered in Section V. The paper ends with a summary of our main results.
\section{Multipolar tidal distortion of the ``shape'' of a BH under the influence of faraway masses}
We wish to describe a physical situation where a BH of mass $M$ is immersed in a generic, stationary, axisymmetric tidal field generated by faraway sources. For simplicity, we shall only consider here the \textit{static} (nonrotating) case. This situation can be described by the Weyl class of static axisymmetric vacuum solutions of Einstein's equations, see, e.g., \cite{Stephani:2003tm} and references therein. The line element of a Weyl metric, in Weyl coordinates, reads
\begin{equation}\label{lineelement}
ds^2=-e^{2\psi}dt^2+e^{2(\gamma-\psi)}\left(d\rho^2+dz^2\right)+\rho^2e^{-2\psi}d\phi^2.
\end{equation}
Einstein's vacuum equations then imply the following equations for the functions $\psi=\psi(\rho,z)$ and $\gamma=\gamma(\rho,z)$ 
\begin{subequations}\label{einstcyl}
\begin{align}
&\f{\partial^2\psi}{\partial\rho^2}+\f{1}{\rho}\f{\partial\psi}{\partial\rho}+\f{\partial^2\psi}{\partial z^2}=0\label{laplcyl},\\
&\f{\partial\gamma}{\partial\rho}=\rho\left[\left(\f{\partial\psi}{\partial\rho}\right)^2-\left(\f{\partial\psi}{\partial z}\right)^2\right],\\
&\f{\partial\gamma}{\partial z}=2\rho\f{\partial\psi}{\partial\rho}\f{\partial\psi}{\partial z}.
\end{align}
\end{subequations}
The linear differential equation (\ref{laplcyl}) is identical to the 3-dimensional axisymmetric ($\phi$-independent) Laplace equation in cylindrical coordinates. Therefore, we can think of $\psi$ as being (minus) the Newtonian potential $U=\sum Gm/r$ generated by an ensemble of axisymmetric bodies: $\psi\equiv-U$. Without loss of generality, the potential $U$ can be decomposed as $U=U_M+U_{\rm ext}$, where $U_M$ refers to the BH of mass $M$ that we consider, and $U_{\rm ext}$ accounts for the external contribution(s). We recall that $U_M$ (for a Schwarzschild BH) corresponds, in Weyl coordinates, to the Newtonian potential generated by a ``rod'', along the $z$ axis, of linear mass density $\mu=1/(2G)$ and of coordinate length $\Delta z=2GM$. The two remaining nonlinear field equations for the metric variable $\gamma$ can be solved by means of a line integral. The superposition of two or more axisymmetric bodies implies the presence of interaction terms in the function $\gamma$: $\gamma=\gamma(U_M)+\gamma(U_{\rm ext})+\gamma_{\rm int}(U_M,U_{\rm ext})$, where $\gamma_{\rm int}$ is bilinear (and nonlocal) in $U_M$ and $U_{\rm ext}$. We shall consider situations where the ``central'' BH of mass $M$ is in ``equilibrium'' within $U_{\rm ext}$, i.e. where the ``elementary flatness'' condition $\lim_{\rho\rightarrow0}\gamma=0$ is satisfied along the portions of the $z$ axis that touch the BH horizon. [As is well-known, a nonvanishing $\gamma_0\equiv\lim_{\rho\rightarrow0}\gamma$ can be interpreted as the presence of a supporting strut or string]. This condition implies the constraint that $U_{\rm ext}$ takes the same value at the north pole ($z=+GM$) than at the south pole ($z=-GM$) of the BH \cite{isrkha64,Gibbons:1974zd,Geroch:1982bv}:
\be\label{NP=SP}
U_{\rm ext}^{NP}=U_{\rm ext}^{SP}\equiv u.
\ee
The notation $u$ for the common value of $U_{\rm ext}$ at the north and south poles is introduced here for later convenience.\\ 
Since the Newtonian potential $U$ obeys a three-dimensional $\phi$ independent Laplace equation, $U_{\rm ext}$ can be decomposed in an axial-multipole expansion in spherical coordinates as 
\begin{equation}\label{multexp}
U_{\rm ext}=\sum_{l=0}^{\infty} U_l=\sum_{l=0}^{\infty} T_l r^l_W P_l(\cos\theta_W).
\end{equation}
Here, ($r_W, \theta_W,\phi$) denote the spherical coordinates associated (as if one were in flat space) to the Weyl coordinates ($\rho,z,\phi$), i.e. $\rho=r_W \sin(\theta_W)$, $z=r_W\cos(\theta_W)$.\\
The coefficients $T_l$ in Eq. (\ref{multexp}) measure the strength of the $l$-th multipolar tidal field within which the considered BH is immersed.
We are aware of the fact that the definition of the $T_l$'s used here heavily relies on the quasilinear properties of the Weyl solutions. We think, however, that, in the \textit{linear} approximation to tidal effects, i.e. in the formal limit where $T_l\rightarrow0$, for all $l$'s, the $T_l$'s become unambiguously defined (and gauge-invariant, as they can be read off at $r\gg GM$). In general, $U_{\rm ext}$ (and therefore each $T_l$) can be thought of as containing the superposition of the tidal fields due to a stationary ensemble of faraway (axisymmetrically distributed) masses. A particularly simple configuration is that where $U_{\rm ext}$ is generated by a single BH of mass $m$, located along the $z$ axis at $z=D\gg GM$. [In this limit, the effect of any ``string'' supporting the perturbing mass $m$ (and the BH) becomes negligible \cite{Suen:1988kq}]. As $D\gg GM$, there exists a wide region (with $GM\ll r_W\ll D$) where space is nearly flat and where we can expand
\be
U_{\rm ext}=\f{Gm}{\sqrt{r_W^2+D^2-2r_WD\cos\theta_W}}
\ee
as
\be\label{uext}
U_{\rm ext}=\sum_{l=0}^{\infty} \frac{Gm}{D^{l+1}} r_W^l P_l(\cos\theta_W).
\ee
This shows that, in this case, the normalization of the tidal coefficients $T_l$ is such that
\be\label{dl+1}
T_l(m)=\f{Gm}{D^{l+1}}.
\ee 
Let us also indicate (from \cite{isrkha64}) the structure of the Weyl solution describing a ``central'' BH of mass $M$, tidally distorted by several BH's of mass $m_i$, described by rods of density $1/(2G)$ located in the intervals [$b_i-Gm_i,b_i+Gm_i$] along the $z$ axis. The solution is described by
\begin{subequations}\label{psiegamma}
\begin{align}
&\psi=\psi_M+\psi_{\rm ext}=\psi_M+\sum_i\psi_{m_i}\\
&\gamma=\gamma_M+\gamma_{\rm ext}=\gamma_M+\sum_i\gamma_{m_i}+2\sum_{i}\gamma_{Mm_i}+\sum_{i,j}\gamma_{m_im_j}+C,\label{gamma}
\end{align}
\end{subequations}
where (using henceforth $G=1$ for simplicity)
\begin{subequations}
\begin{align}
&\psi_M=\f{1}{2}\ln\left[\f{R_M^++R_M^--2M}{R_M^++R_M^-+2M}\right],\label{psiemme}\\
&\gamma_M=\f{1}{2}\ln\left[\f{(R_M^++R_M^-)^2-4M^2}{4R_M^+R_M^-}\right],\label{gammaemme}
\end{align}
\end{subequations}
and where each $m_i$-dependent contribution reads (when suppressing the index $i$ for readability)
\begin{subequations}
\begin{align}
&\psi_m=\f{1}{2}\ln\left[\f{R_m^++R_m^--2m}{R_m^++R_m^-+2m}\right],\label{psim}\\
&\gamma_m=\f{1}{2}\ln\left[\f{(R_m^++R_m^-)^2-4m^2}{4R_m^+R_m^-}\right],\\
&\gamma_{Mm}=\f{1}{4}\ln\left[\f{E_{(M+,m-)}E_{(M-,m+)}}{E_{(M+,m+)}E_{(M-,m-)}}\right],
\end{align}
\end{subequations}
with $\gamma_{m_im_j}$ given by a similar expression (obtained by replacing $M\rightarrow m_i$, $m\rightarrow m_j$).\\
Here,
\begin{subequations}\label{def}
\begin{align}
&R_M^\pm=\sqrt{\rho^2+(Z_M^\pm)^2},\qquad R_m^\pm=\sqrt{\rho^2+(Z_m^\pm)^2},\\
&Z_M^\pm=z\pm M,\qquad Z_m^\pm=z-(b\mp m),\\
&E_{(M\pm,m\pm)}=\rho^2+R_M^\pm R_m^\pm+Z_M^\pm Z_m^\pm.
\end{align}
\end{subequations}
In Eq. (\ref{gamma}), $C$ denotes an integration constant which must be chosen (after having satisfied the condition (\ref{NP=SP})) so that $\lim_{\rho\rightarrow0}\gamma$ vanishes on the portions of the $z$ axis which touch the ``central'' BH. When the condition (\ref{NP=SP}) is satisfied, and the constant $C$ in (\ref{gamma}) is suitably chosen, it has been shown \cite{isrkha64,Gibbons:1974zd,Geroch:1982bv} that the locus $\rho=0$, $z\in[-M,M]$, is a smooth (Killing) horizon. To see that its structure is that of a distorted BH, one would need to replace the Weyl coordinates ($\rho$, $z$, $\phi$) first by Schwarzschild-like coordinates ($R$, $\theta_S$, $\phi$), such that
\begin{subequations}\label{slike}
\begin{align}
&\rho=\sqrt{R^2-2MR}\sin\theta_S,\\
&z=(R-M)\cos\theta_S,
\end{align}
\end{subequations} 
and then by (horizon-regular) Kruskal (or Eddington-Finkelstein) coordinates. For our present purpose, which is to read off the distorted geometry of the horizon $R=2M$ ($\rho=0$), it will be enough to consider the geometry (\ref{lineelement}) in Schwarzschild-like coordinates, i.e. using the following consequences of Eqs. (\ref{psiemme}), (\ref{gammaemme}), (\ref{def}) and (\ref{slike})	
\begin{subequations}
\begin{align}
&e^{2\psi_M}=1-\f{2M}{R},\\
&e^{2(\gamma_M-\psi_M)}=[(R-M)^2-M^2\cos^2\theta_S]^{-1},\\
&ds^2=-e^{2\psi_{\rm ext}}\left(1-\f{2M}{R}\right)dt^2+e^{2\left(\gamma_{\rm ext}-\psi_{\rm ext}\right)}\left[\f{dR^2}{1-\f{2M}{R}}+R^2d\theta_S^2\right]+e^{-2\psi_{\rm ext}}R^2\sin^2\theta_Sd\phi^2.\label{geometry}
\end{align}
\end{subequations} 
Here, as indicated in the particular case of (\ref{psiegamma}), $\psi_{\rm ext}$ and $\gamma_{\rm ext}$ are defined such that $\psi=\psi_M+\psi_{\rm ext}$, $\gamma=\gamma_M+\gamma_{\rm ext}$, where $\psi_M$ and $\gamma_M$ are the undistorted ``Schwarzschild'' values (\ref{psiemme}) and (\ref{gammaemme}).\\
It is easily seen that Eqs. (\ref{einstcyl}) imply that the following equality
\be\label{kappa}
\left[\gamma_{\rm ext}-2\psi_{\rm ext}\right]_{\rho=0}=const\equiv2u
\ee
holds on the horizon ($\rho\rightarrow0$). Here, $u$ is, as in Eq. (\ref{NP=SP}), the common value of $U_{\rm ext}\equiv-\psi_{\rm ext}$ at the north and south poles. Restricting the geometry (\ref{geometry}) to $t=t_0=const$ and $R=R_0=const$ (and then taking the limit $R_0\rightarrow2M$), and using Eq. (\ref{kappa}), allows us to read off the geometry induced on the horizon:
\be\label{ds2hor}
ds^2|_{R=2M}=(2Me^u)^2\left[\f{1}{\beta(\mu)}d\mu^2+\beta(\mu)d\phi^2\right],
\ee
where we have introduced the convenient variable
\be\label{muesse}
\mu\equiv\cos\theta_S=\left(\f{z}{M}\right)_{\rho=0},\qquad (-1\le\mu\le+1),
\ee
and where the function $\beta(\mu)$ describing the distorted horizon geometry is given by
\begin{subequations}
\begin{align}
&\beta(\mu)=(1-\mu^2)\hat{\beta}(\mu),\label{beta}\\
&\hat{\beta}(\mu)=e^{2\left(\bar{U}_{\rm ext}(\mu)-u\right)}.\label{betahat}
\end{align}
\end{subequations}
Here $\bar{U}_{\rm ext}(\mu)=-\psi_{\rm ext}|_{\rho=0}$ is the value of $U_{\rm ext}$ along the horizon, and $u$ is, as in Eq. (\ref{NP=SP}), the common value of $U_{\rm ext}$ at the north and south poles.\\
Note that the prefactor giving the overall length scale of the horizon geometry is not the mass parameter $M$ entering the Weyl metric, but rather the ``blueshifted'' mass parameter
\be\label{mu}
M_u=Me^u.
\ee
While the mass parameter $M$ is equal \cite{Geroch:1982bv} to the \textit{Komar} mass of the central BH, Eq. (\ref{ds2hor}) shows that the \textit{irreducible} mass \cite{Christodoulou:1970wf,Christodoulou:1972kt} of the central BH is
\be
M_{\rm irr}\equiv\sqrt{\f{A}{16\pi G^2}}=M_u,
\ee
where $A$ denotes the area of the horizon. If we were to consider a physical process where our ensemble of gravitationally interacting masses would change their relative positions in an adiabatic manner, we should be careful to maintain all the irreducible masses (i.e. all the individual BH entropies) \textit{fixed} during the process. This would mean that the Komar mass parameters $M$, $m_i$ would need to change as the relative positions change, so as to compensate the variation of the blueshift factor $e^u$ in Eq. (\ref{mu}). However, as our purpose is to study the \textit{fractional} distortion of the horizon geometry associated to external tidal fields, we will, in the following, factor out the overall scale factor $(2M_u)^2$ in Eq. (\ref{ds2hor}) and focus on the shape of the $\beta(\mu)$-dependent conformal geometry defined by the square bracket on the right-hand side of Eq. (\ref{ds2hor}).\\
\\
It can be easily proven that the Gauss curvature $K=\tfrac{1}{2}R^{(2)}=\tfrac{1}{2}R^{\alpha\beta}_{\quad\alpha\beta}$ of the distorted horizon geometry (\ref{ds2hor}) is given in terms of the second $\mu$-derivative of the function $\beta(\mu)$ by
\be\label{capitalcappa}
(2M_u)^2K=-\frac{1}{2}\beta^{''}(\mu).
\ee
As a check on Eq. (\ref{capitalcappa}), one can verify the Gauss-Bonnet theorem:
\be\label{gbt}
\int\int KdA=\int\int d\mu d\phi\left[-\f{1}{2}\beta^{''}\right]=-\pi\left[\beta^{'}(\mu)\right]_{-1}^{+1}=2\pi\left[\hat{\beta}(+1)+\hat{\beta}(-1)\right].
\ee
When the horizon is regular, i.e. when Eq. (\ref{NP=SP}) is satisfied so that $\hat{\beta}^{NP}=\hat{\beta}^{SP}=1$, Eq. (\ref{gbt}) yields $\int\int KdA=4\pi$ as necessary for a horizon having the topology of a $2$-sphere.\\
\\
Let us now insert in Eq. (\ref{capitalcappa}) our general parametrization (\ref{multexp}) of an external tidal field by the sequence of tidal coefficients $T_l$. To do so, we must replace the Weyl-associated ``spherical coordinates'' $r_W$, $\theta_W$ entering Eq. (\ref{multexp}) in terms of $\rho$, $z$, before taking the (singular) horizon limit $\rho\rightarrow0$, $z\rightarrow M\mu$ which defines the horizon value of $U_{\rm ext}$ entering the definition of the function $\beta(\mu)$, Eq. (\ref{beta}). For instance, the $l=2$ term in Eq. (\ref{multexp}) would be rewritten as
\be
r^2_WP_2(\cos\theta_W)=r^2_W\f{3\cos^2\theta_W-1}{2}=\frac{1}{2}\left(3z^2-\left(\rho^2+z^2\right)\right)=z^2-\tfrac{1}{2}\rho^2,
\ee
whose horizon limit is simply $z^2=(M\mu)^2$. More generally, it is easy to see (using $P_l(1)=1$) that the horizon limit of $r_W^lP_l(\cos\theta_W)$ is simply $z^l=(M\mu)^l$, so that
\be
\bar{U}_{\rm ext}(\mu)=U_{\rm ext}|_{\rm horizon}=\sum_lT_lM^l\mu^l.
\ee 
Subtracting the value of $U_{\rm ext}|_{\rm horizon}$ at the north pole $\mu=+1$ finally yields the following explicit link between the Gauss curvature of the horizon and the sequence of external tidal coefficients $T_l$:
\be
(2M_u)^2K(\mu)=-\f{1}{2}\f{\partial^2}{\partial\mu^2}\left[(1-\mu^2)e^{2\sum_lT_lM^l(\mu^l-1)}\right].
\ee
At this stage, it is natural to decompose also the (scaled) Gauss curvature in multipolar components, say
\be
(2M_u)^2K(\mu)\equiv\sum_lc_lP_l(\mu).
\ee
Before deriving a formula giving the coefficients $c_l$ of the multipolar expansion of the Gauss curvature, let us note the two general, exact results
\begin{subequations}
\begin{align}
&c_0=1,\label{c0}\\
&c_1=0.\label{c1}
\end{align}
\end{subequations}
Eq. (\ref{c0}) is a restatement of the Gauss-Bonnet theorem checked above. As for the result (\ref{c1}), it follows from the general structure (\ref{capitalcappa}) with a function $\beta(\mu)$ satisfying (\ref{beta}) and (\ref{betahat}). Indeed, using $P_1(\mu)=\mu$ and the orthogonality of the Legendre polynomials, $c_1$ is given by
\be
c_1=-\f{3}{4}\int_{-1}^{+1}d\mu \mu\beta^{''}(\mu)=-\f{3}{4}\left[\mu\beta^{'}-\beta\right]_{-1}^{+1}.
\ee
Using the expression (\ref{beta}), where $\hat{\beta}$ takes the value $1$ at both endpoints, on easily finds that $c_1$ necessarily vanishes.\\
The orthogonality properties of the Legendre polynomials, $\int_{-1}^{+1}P_l(\mu)P_{l^{'}}(\mu)=2\delta_{ll^{'}}/(2l+1)$, then yield the following expression for the general $c_l$'s as functions of the sequence of the $T_l$'s:
\be\label{clcomplete1}
c_l=-\f{2l+1}{4}\int_{-1}^{+1} d\mu P_l(\mu)\f{\partial^2}{\partial\mu^2}\left[(1-\mu^2)e^{2\sum_{l'}T_{l'}M^{l'}(\mu^{l'}-1)}\right].
\ee
Note that $c_l$ is a \textit{nonlinear} function of \textit{all} the $T_{l'}$'s, with $l'$ being a priori unrelated to $l$. To better understand this function, we can however expand the exponential in Eq. (\ref{clcomplete1}) in powers of the $T_l$'s. In so doing, and in using the orthogonality of $P_l(\mu)$ to all the powers $\mu^{l'}$ when $l'<l$, one finds that $c_l$ is the sum of a ``principal'' contribution $\propto T_l$, plus two types of extra contributions: (i) linear in $T_{l'}$,  which arise only with $l'=l+2, l+4, l+6$ etc; and (ii) nonlinear in $T_{l'}$, which can be quadratic $\propto T^2$, cubic $\propto T^3$, etc. Summarizing, we have the structure (for $l\geq2$)
\be\label{clcomplete}
c_l= n_lT_lM^l+n_{l,2}T_{l+2}M^{l+2}+n_{l,4}T_{l+4}M^{l+4}+...+\sum_{l^{'}l^{''}}n_{ll^{'}l^{''}}T_{l^{'}}T_{l^{''}}M^{l^{'}+l^{''}}+...\quad,
\ee
where the $n_l$'s, etc., denote some numerical coefficients.\\
In the ``linear'' approximation to tidal effects (i.e. in the formal limit $T_{l'}\rightarrow0$) we can neglect the nonlinear contributions $\propto T^2, T^3$ etc. In addition, if we focus, for simplicity, on the case where the (minimum) distance $D$ to the external masses creating $U_{\rm ext}$ becomes large, we find, in view of $T_l\sim1/D^{l+1}$, Eq. (\ref{dl+1}), that the contributions proportional to $T_{l+2n}$ are parametrically smaller than the ``principal'' contribution $\propto T_l$ by a factor $(M/D)^{2n}$ which tends to zero. In this double limit, we conclude that a kind of Hooke's law is valid in that the $l$-th multipolar component of the horizon curvature, $c_l$, is proportional to the $l$-th tidal strength coefficient $T_l$, i.e. 
\be
c_l\simeq n_lT_lM^l,
\ee  
where
\begin{equation}
\begin{aligned}
&n_l=-\f{(2l+1)}{2}\int_{-1}^{+1} d\mu P_l(\mu) \f{\partial^2}{\partial\mu^2}\left[(1-\mu^2)(\mu^l-1)\right]=\\
&=\f{(2l+1)}{2}(l+2)(l+1)\int_{-1}^{+1} d\mu P_l(\mu)\mu^l=(l+2)(l+1)\f{2^l(l!)^2}{(2l)!}.
\end{aligned}
\end{equation}
Finally, we conclude that, in the leading approximation, the Gauss curvature induced by a general external tidal potential reads (remembering the exact results of (\ref{c0}) and (\ref{c1}))
\be\label{pag10}
(2M_u)^2K(\mu)\simeq 1+\sum_{l=2}^{\infty}(l+2)(l+1)\f{2^l(l!)^2}{(2l)!}M^lT_lP_l(\mu).
\ee
Let us now convert this result in terms of the ``shape'' Love number $h_l$. We define this number, in the present general relativistic setting, by paralleling the definition Love used in Newtonian gravity \cite{love1911}. The basic idea is to write, for each multipole $l$, the specific potential energy $g_0(\delta R)_l$ associated to the $l$-th ``tidal bulge'' of height $(\delta R)_l$ as $h_l$ times the \textit{external} tidal potential, evaluated (as it would be in Newtonian gravity) at the undisturbed radius of the considered object , i.e.
\be\label{gzerodeltar}
g_0(\delta R)_l=h_lU^{\rm ext}_l(R_0),
\ee
or, equivalently, 
\be\label{deltaR/R}
\left(\f{\delta R}{R_0}\right)_l=h_l\f{U_l^{\rm ext}(R_0)}{g_0R_0}.
\ee
Here, $g_0=GM/R_0^2$ denotes the (Newtonian) surface gravity of the considered object. We recall that the same (Newtonian-looking) formula is also used in BH theory to define the (renormalized) surface gravity of a Schwarzschild BH, if, as we shall do, one defines the ``radius'' of a BH as its (areal) radius $R_0=2GM$. [In view of the remarks above, and of our focus on the linear approximation to tidal effects, we shall not worry here about the possible distinction between the Komar mass $M$ and the irreducible mass $M_u=Me^u\simeq M(1+u+\mathcal{O}(u^2))$.]. As for $U_{\rm ext}(R_0)$ we define it, as it would be done in Newtonian gravity, by taking the analytic continuation down to $R=R_0$ of the $l$-th multipolar order \textit{asymptotic} Newtonian potential. In the (intermediate) asymptotic domain $R_0\ll R\ll D$ the external tidal potential (\ref{multexp}) can be written in Schwarzschild-type coordinates ($R, \theta_S$), Eqs. (\ref{slike}), as $U_{\rm ext}\simeq\sum_lT_lR^lP_l(\cos\theta_S)$. Formally continuing this Newtonian-like expression down to $R=R_0$, with fixed Schwarzschild-like colatitude $\theta_S$, leads to 
\be\label{U/R}
U_{\rm ext}(R_0,\mu)=\sum_lU_l^{\rm ext}(R_0)P_l(\mu)=\sum_lT_lR_0^lP_l(\mu),
\ee
where, in agreement with Eq. (\ref{mu}), we have identified $\cos\theta_S$ with the variable $\mu$ used in the study of the horizon geometry above.\\
Summarizing, the ``shape'' (or ``height'') Love numbers $h_l$ are defined by writing
\be\label{defh}
\left(\f{\delta R}{R_0}\right)_l=h_l\f{T_lR_0^l}{g_0R_0}=h_l\f{T_lR_0^{l+1}}{GM}.
\ee
In order to compare (\ref{defh}) to our result Eq. (\ref{pag10}) above, we need to convert a general multipolar expansion of the Gauss curvature, say,
\be\label{kk}
R_0^2K=1+\sum_{l\ge2}c_lP_l(\mu)
\ee
into a corresponding ``height'' expansion, say (denoting $\epsilon_l\equiv(\delta R/R_0)_l$)
\be\label{rofmu}
\f{\delta R(\mu)}{R_0}=\sum_l\epsilon_lP_l(\mu).
\ee
This is naturally done by defining $\epsilon_l$ so that the Gauss curvature on the $2$-surface $\mathbf{X}(\theta,\phi)=(R_0+\delta R(\mu))\bf{N}(\theta,\phi)$ (with $N^1=\sin\theta\cos\phi$, $N^2=\sin\theta\sin\phi$, $N^3=\cos\theta$) embedded in an auxiliary $3$-dimensional Euclidean space, coincides with Eq. (\ref{kk}). A straightforward calculation shows that this coincidence is obtained (in first order in the perturbation away from a round sphere) if
\be\label{cl}
c_l=[l(l+1)-2]\epsilon_l\equiv(l-1)(l+2)\epsilon_l.
\ee
Note in passing that any eventual $l=1$ contribution $\epsilon_1$ in (\ref{rofmu}) is canceled in the corresponding curvature expansion. This agrees with the general result (\ref{c1}) above, and allows us to consider only multipolar orders $l\ge2$.\\
Finally, by combining Eqs. (\ref{defh}),(\ref{deltaR/R}), (\ref{U/R}) and (\ref{cl}), we find that the $l$-th shape Love number of a BH is given by 
\begin{equation}\label{hl}
h_l=\frac{l+1}{l-1}\frac{(l!)^2}{2(2l)!}.
\end{equation}
In particular, we find for the first values of $l$:
\be
h_2=\f{1}{4}, \qquad h_3=\f{1}{20}, \qquad h_4=\f{1}{84}.
\ee
We see that $h_l$ diminishes rapidly as $l$ increases.\\
\\
Note that all these Love numbers are smaller than $1$. This contrasts with the Love numbers of a  perfect-fluid star in Newtonian gravity which are given in terms of their ``second'' Love numbers $k_l$ by $h_l^{\rm Newton}=1+2k_l^{\rm Newton}$ \cite{Damour:2009vw}, so that $h_l^{\rm Newton}>1$ (because $k_l^{\rm Newton}>0$). In the recent work \cite{Damour:2009vw}, it was, however, found that the shape Love numbers $h_l^{NS}$ of a general relativistic fluid star (say a neutron star) have a strong dependence on the self-gravity of the star, and that $h_l^{NS}$ continuously decrease, as the ``compactness parameter'' $c=GM/R$ increases, from the Newtonian value $h_l^{\rm Newton}>1$ down to a value close to the above black-hole values, $h_l^{BH}$, as the compactness reaches its maximum possible value. It was also found that the function $h_l^{NS}(c)$ becomes equal to $h_l^{BH}$ as the compactness parameter $c$ is formally continued to the black-hole value $c^{BH}=1/2$.\\
\\
Using the Stirling approximation, one finds that, for large values of $l$,
\be
h_l\simeq \f{\sqrt{2\pi l}}{2^{2l+1}},
\ee
so that $h_l$ essentially decreases as $4^{-l}$. This rather fast decrease of $h_l$, as $l$ increases, has useful practical consequences. Indeed, in the physically most relevant case where one is interested in the tides raised on a certain BH of mass $M$ by another BH (of mass $m$), located\footnote{Note the that extra linear terms $n_{l,2} T_{l+2}M^{l+2}+...$ that we neglected in Eq. (\ref{clcomplete}) are, in order of magnitude, essentially equivalent to correcting the distance $D$ entering $T_l\propto D^{-l-1}$ in the leading term $n_lT_lM^l$ by $D\rightarrow D(1+c_2M^2/D^2+c_4M^4/D^4)+...$. Such corrections include, in particular, possible effects of coordinate transformations on $D$.} at a distance $D$, we see that the tidal expansion, Eq. (\ref{uext}), of $U_{\rm ext}$ (which, from a Newtonian point of view, would be expected to converge on the horizon as a geometric series $\sum_l(R_0/D)^l$) raises on the horizon a tide whose height converges like
\be
\f{\delta R}{R_0}\sim\sum_{l=2}^{\infty}\f{m}{M}\sqrt{l}\left(\f{R_0}{4D}\right)^{l+1}P_l,
\ee
which is essentially a geometric series $\sim\sum_l(\tfrac{R_0}{4D})^l$. The appearance of $\tfrac{R_0}{4D}$ instead of $\tfrac{R_0}{D}$ implies a rather fast convergence, even when $D$ is not much larger than $R_0$. Physically, this means that taking into account only the quadrupolar tide $\propto h_2$ will probably suffice to give a good estimate of the full tide, even when the companion BH is rather near. In a later section, we shall, however, see that this conclusion no longer holds when $m\ll M$ and when the mass $m$ is allowed to come very near the horizon of $M$. By contrast, when $m\sim M$, the finite size of the horizon of $m$ does not allow the distance $D$ to become smaller than some minimum value $D_{\rm min}\propto M+m$, so that one might hope, in view of the remarks above, that the (linear) quadrupolar tide alone might give a good estimate of the tidal deformations of both holes down to the point where they formally touch. In the case of the quadrupolar tide, one knows from other studies  \cite{manasse2,Poisson:2005pi} that a good measure of the $l=2$ tidal field $T_2$ is given by some component of the Riemann (or Weyl) tensor, which, in the exterior of a Schwarzschild BH of mass $m$ is $\propto m/R^3$, where $R$ is the usual (areal) Schwarzschild coordinate (rather, say, than the Weyl radial coordinate distance $D$ that we have been using up to now.). This leads one to expect that, when $m$ and $M$ are comparable, $m$ will raise on $M$ a tide of height 
\begin{equation}\nonumber 
\left(\f{\delta R}{R_0}\right)_M\simeq h_2\frac{m}{R^3}\frac{R_0^3}{M}P_2\sim\frac{1}{4}\frac{m}{M}\left(\frac{R_0(M)}{R}\right)^3.
\end{equation} 
Upon ``contact'', i.e. when $R=R_0(M)+R_0(m)=2(M+m)$, we get $(\delta R/R_0)_M\sim\tfrac{1}{4}\tfrac{mM^2}{(m+M)^3}$. This result would predict a maximum deformation $\delta R/R_0\sim1/27$, reached when $M=2m$. We note that this deformation is surprisingly small. One should, however, remember that its ``derivation'' relied on several approximations that cease to be valid in the limit we consider.\\
We have tried to further study the tidal deformations of comparable-mass BH's by considering the multi-black-hole Weyl solutions recalled in Eqs. (\ref{psiegamma}) and (\ref{def}) above. The problem, however, is that these solutions do not allow one to study, in a physically relevant way, a process where two nearly isolated BH's get very near each other. Indeed, the total configuration will always include some struts or strings, that are necessary for global equilibrium. One can arrange two BH's, among $N$, to be free of such strings. However, when these two ``central'' BH's get close to each other, one will need other BH's (supported by strings) to become also very close to the central BH's. This generates additional strong forces and accelerations (related to $l=1$ tidal terms) that mess up the pure $l=2$ tidal heights coming from the (free fall) gravitational interaction of the two central holes. As a consequence, we could not use Weyl solutions to check the above prediction that ``touching'' BH's are only slightly deformed.\\
Instead, we could, however, use well separated, multi-black-hole Weyl solutions to check the values of the first few Love numbers. For instance, one can take an asymmetric $3$-black-hole Weyl solution, made of a central BH of mass $M$ (``located'' at $z=0$, i.e. corresponding to a ``rod'' in the interval $z\in[-M,M]$), and of two ``satellite'' BH's: one of mass $m_1$ located at $z=b_1>0$, and one of mass $m_2$ located at $z=-b_2<0$. One then considers a limit where $b_1\sim b_2$ tend to infinity (e.g. keeping $m_1\sim m_2$ finite). In this limit, one can expand the solution (\ref{psiegamma}) and (\ref{def}) in inverse powers of $b_1$ and $b_2$. One finds that, by using a suitable constant $C$ in Eq. (\ref{gamma}), and by tuning $m_2$ (given $m_1$, $b_1$ and $b_2$) so that
\begin{equation}
\frac{m_1}{b_1^2}+M^2\frac{m_1}{b_1^4}=\frac{m_2}{b_2^2}+M^2\frac{m_2}{b_2^4}+\mathcal{O}\left(\f{1}{b^5}\right),
\end{equation}
one can arrange to have the central BH of mass $M$ to hold in equilibrium without supporting struts (i.e. with $\lim_{\rho\rightarrow0}\gamma=0$ on both sides of the $z$ axis, and therefore with a smooth horizon and $\hat{\beta}^{NP}=\hat{\beta}^{SP}=1$). Then, we find that the Gauss curvature of the central horizon reads
\begin{equation}
(2M_u)^2K=1+8M^2\left(\frac{m_1}{b_1^3}+\frac{m_2}{b_2^3}\right)P_2+8M^3\left(\frac{m_1}{b_1^4}-\frac{m_2}{b_2^4}\right)P_3+\mathcal{O}\left(\f{1}{b^5}\right).
\end{equation}
Using the definition above of the Love numbers, one easily checks that this result is equivalent to $h_2=1/4$ and $h_3=1/20$, in agreement with our general formula above.
\section{Electromagnetic case: influence of faraway charges on the surface charge density of a BH}
As a contrast to the gravitational polarizability properties of BH's, let us now consider their electric polarizability properties. For simplicity, we consider a physical situation where an uncharged BH of mass $M$ is immersed in a general electric field generated by a static axisymmetric configuration of faraway charges.\\   
The background metric of the Schwarzschild black hole is
\begin{equation}\label{Schwarzschild}
ds^2=-\left(1-\frac{2M}{R}\right)dt^2+\frac{1}{1-\frac{2M}{R}}dR^2+R^2\left(d\theta^2+\sin^2\theta d\phi^2\right).
\end{equation}
In the linearized approximation, the electromagnetic field $F_{\mu\nu}=\partial_\mu A_\nu-\partial_\nu A_\mu$ generated by the faraway charges must satisfy (outside the location of the charges)
\be\label{24}
F^{\mu\nu}_{\quad;\nu}=0,
\ee
where the semicolon denotes the covariant derivative. Eq. (\ref{24}) yields a second-order partial differential equation for the scalar potential $A_0=-V$. [Here $V$ denotes the usual electric potential such that the electric field $\bf{E}=-$ \mbox{\boldmath$\nabla$} $ V=+$ \mbox{\boldmath$\nabla$} $ A_0$.] Assuming that all the charges that generate $A_0$ are at a distance $D\gg 2M$, we can, as in the gravitational case, consider that, in the intermediate domain $2M\ll R\ll D$ (where the spacetime is approximately flat), the potential $A_0$ admits a flat-space multipolar expansion of the general type
\be\label{Vasympt}
A_0^{\rm asympt}(R,\theta)=\sum_{l=0}^{\infty}\tau_lR^lP_l(\cos\theta).
\ee
Here the coefficients $\tau_l\sim +\partial^lA_0\sim+\partial^{l-1}E$ are the electric analogs of the tidal coefficients $T_l\sim+\partial^lU\sim+\partial^{l-1}g$ that entered the asymptotic tidal expansion of the external gravitational potential $U_{\rm ext}$, Eq. (\ref{multexp}). The sign convention is chosen so that a uniform electric field $\mathbf{E}=E\mathbf{e}_z$ directed along the positive $z$ axis corresponds to $\tau_1=+E$. In addition, the normalization of the $\tau_l$'s is such that, in the particular case where the electric field that we consider is generated by a pointlike test charge $q$ located at a large distance $D$ along the $z$ axis (in a flat region), $\tau_l$ is simply equal to
\be
\tau_l=-\frac{q}{D^{l+1}},
\ee
so that the asymptotic scalar potential reads 
\be
A_0^{\rm asympt}=-V_{\rm asympt}=-\f{q}{\sqrt{r^2+D^2-2rD\cos\theta}}=-\sum_{l=0}^{\infty}\frac{q}{D^{l+1}}r^lP_l(\cos\theta).
\ee
Let us now take into account the effect of the background  spacetime curvature, Eq. (\ref{Schwarzschild}). Because of the spherical symmetry of the background, and the assumed axisymmetry, one can decompose the exact $A_0=-V$ in a series of Legendre polynomials, say
\be\label{A0}
A_0=\sum_{l=0}^{\infty}a_l(R)P_l(\cos\theta)\equiv\sum_{l=0}^{\infty}\sqrt{1-\f{2M}{R}}w_l(R)P_l(\cos\theta),
\ee
where the functions $w_l(R)\equiv a_l(R)/\sqrt{1-2M/R}$ can be shown to obey the differential equation
\be\label{legendrelike}
\left(1-\bar{R}^2\right)\f{d^2w_l}{d\bar{R}^2}-2\bar{R}\f{dw_l}{d\bar{R}}+\left[l(l+1)-\f{1}{1-\bar{R}^2}\right]w_l=0
\ee
where $\bar{R}=R/M-1$. Following \cite{Bini:2006dp}, we remark that Eq. (\ref{legendrelike}) has the form of a general Legendre equation ($m=1$), and admits two independent solutions, which can be expressed in terms of the associated Legendre functions of the first and second kind, $P_l^1(\bar{R})$ and $Q_l^1(\bar{R})$, respectively. As a result, the expansion (\ref{A0}) contains two classes of radial functions $a_l(R)$: one class of solutions, say $g_l$, is regular on the horizon, and behaves like $R^l$ for $R\rightarrow\infty$, while the other, say $f_l$, is logarithmically singular (except when $l=0$) on the horizon and behaves like $R^{-(l+1)}$ for $R\rightarrow\infty$. We are interested here in the unique radial solution $a_l(R)$ that is regular on the horizon and grows like $R^{l}$ when $R\rightarrow\infty$, i.e. the $g_l$'s, (so that the corresponding term in Eq. (\ref{A0}) can match the term $\propto \tau_lR^l$ in Eq. (\ref{Vasympt})). Such a solution is given by $g_0=1$ when $l=0$, and
\be\label{regular}
g_l(R)=\frac{(2M)^l(l-1)!l!}{(2l)!}(R-2M)\frac{dP_l(R/M-1)}{dR},\ \ {\rm when}\ \ l\neq0,
\ee
where the normalization of (\ref{regular}) is chosen such that $g_l(R)\simeq R^l$ for $R\rightarrow\infty$.
With this normalization, the unique, horizon-regular solution of (\ref{24}) which asymptotically agrees with (\ref{Vasympt}) (and corresponds, when $l=0$, to an uncharged BH) reads
\be
A_0(R,\theta)=\sum_{l=0}^{\infty}\tau_lg_l(R)P_l(\cos\theta).
\ee
Let us now consider what is the ``charge density'' $\sigma$ \cite{Hanni:1973fn} induced on the horizon of the BH by the electromagnetic field, namely
\begin{equation}\label{sigmadef}
\sigma\equiv\frac{1}{4\pi}\left[\f{\partial A_0}{\partial R}\right]_{R=2M}=-\f{1}{4\pi}\left[\f{\partial V}{\partial R}\right]_{R=2M}.
\end{equation}
Decomposing $\sigma$ into multipoles, 
\be
\sigma(\cos\theta)=\sum_{l=1}^{\infty}\sigma_lP_l(\cos\theta),
\ee
and using the identity
\be\label{poldiff}
\left[\f{d}{dx}P_l(x)\right]_{x=1}=\f{l(l+1)}{2},
\ee
one easily finds that (with $R_0=2M$)
\be\label{sigmal}
4\pi R_0\sigma_l=h_l^{EM}\tau_lR_0^l,\qquad l\ge1, 
\ee
where
\be\label{hlem}
h_l^{EM}=\f{l!(l+1)!}{(2l)!},\qquad l\ge1.
\ee
On the left-hand side of Eq. (\ref{sigmal}) we have introduced the quantity $4\pi R_0\sigma_l$ which has the dimension of an electric potential, i.e. the same dimension as the $l$-th multipolar component $\tau_lR_0^l$ of the asymptotic electric potential $A_0$ (formally evaluated for a radius $R=R_0=2M$). Eq. (\ref{sigmal}) is the electric analog of Eq. (\ref{gzerodeltar}): its right-hand side contains the $R\rightarrow R_0$ continuation of a ``tidal'' potential, while its left-hand side contains the ``effect'' of its tidal potential on the horizon (here the induction of a charge density). We can therefore consider that the dimensionless proportionality coefficient $h_l^{EM}$ entering the linear relation (\ref{sigmal}) is the \textit{electric analog} of the $h_l$ Love number (hence our notation).\\
It is interesting to note that this ``electric Love number'' is rather similar to its gravitational analog. It is related to it by the simple factor
\be
h_l^{EM}=2(l-1)h_l.
\ee 
The first few values are $h_1^{EM}=1$, $h_2^{EM}=1/2$, $h_3^{EM}=1/5$. For large values of $l$, $h_l^{EM}$ decays as
\be
h_l^{EM}\sim\f{l\sqrt{2\pi l}}{4^l}.
\ee
As in the gravitational case, we can therefore expect that, as long as the inducing charge $q$ is at a distance $D\gtrsim R_0$, its ``electric tidal'' effect will be dominated by the lowest multipole, i.e., in the present case, by the induction of a \textit{dipolar} charge distribution $4\pi\sigma_1=h_1^{EM}\tau_1$, i.e., (using our result $h_l^{EM}=1$) $4\pi\sigma_1=E^{\rm asympt}$ where $E^{\rm asympt}$ is the \textit{asymptotic} ($R_0\ll R\ll D$) electric field strength.\\
\\
In this respect, it is also interesting to compare the electric Love numbers of a BH to those of a conducting sphere in flat space. We recall that a BH is analogous to a conducting sphere in that, at equilibrium, it is an equipotential surface \cite{Hanni:1973fn}, and that, in a general nonequilibrium situation, it exhibits dissipative properties (Ohm's law, Joule's law) similar to those of a conducting shell with surface resistivity equal to $4\pi=377 \Omega$ \cite{Damour:1978cg,Znajek:1978,Damour:1979,Damour:1982}. The general ``tidal'' $l$-th multipolar component of the electric potential around an uncharged, conducting sphere of radius $R_0$ (in flat space) is easily found to be 
\be
A_0(R,\theta)=\tau_0+\sum_{l=1}^{\infty}\tau_l\left(R^l-\f{R_0^{2l+1}}{R^{l+1}}\right)P_l(\cos\theta).
\ee
Computing the corresponding multipole-expanded charge density $\sigma$, Eq. (\ref{sigmadef}), we then deduce from the definition (\ref{sigmal}), the flat-space values of the electric Love number of a conducting sphere
\be\label{conductingh}
h_l^{\rm sphere}=2l+1.
\ee
Note that these flat-space values are larger than $1$ and they grow with $l$ ($h_1^{\rm sphere}=3$, $h_2^{\rm sphere}=5$). This contrasts with the corresponding BH values ($h_1^{BH}=1$, $h_2^{BH}=1/2$) that decrease with $l$.
\section{Electromagnetic case: near-horizon limit}
Up to now, we have been considering (both in the gravitational case and in the electromagnetic one) a BH immersed in the (gravitational or electromagnetic) ``tidal field'' generated by a configuration of faraway sources. To complete our understanding of tidal effects, it is, however, interesting to consider also the case where a (test) mass $m$, or a (test) charge $q$, gets very close to the horizon. As a prelude to studying the gravitational case, we shall consider in this section the case where an external test electric charge gets very close to the horizon of a Schwarzschild BH. This was the situation studied long ago by Hanni and Ruffini \cite{Hanni:1973fn}.\\
\\
We then consider a test charge $q$ located\footnote{Note that, in this subsection, $D$ denotes a Schwarzschild-coordinate radial distance.} at $z=D$ along the positive $z$ axis ($\theta=0$). Outside the radial location of the charge, the multipolar components $a_l(R)$ of $A_0(R,\theta)$, Eq. (\ref{A0}), satisfy Eq. (\ref{legendrelike}) (after the transformation $a_l(R)=w_l(R)\sqrt{1-2M/R}$). Above, it was enough for our purpose to consider only the asymptotically-growing, horizon-regular solution of (\ref{legendrelike}), for which $a_l(R)$ is given by Eq. (\ref{regular}). Now, we need to consider also the asymptotically-decreasing, horizon-singular solution of (\ref{legendrelike}), for which $a_l(R)$ is given by \cite{Bini:2006dp}
\be\label{flr}
f_l(R)=-\frac{(2l+1)!}{2^l(l+1)!l!M^{l+1}}(R-2M)\frac{dQ_l(\tfrac{R}{M}-1)}{dR},\ \ \forall l\\
\ee
The normalization of $f_l(R)$ has been chosen so that 
\be
f_l(R)\simeq R^{-(l+1)}, \qquad R\rightarrow \infty.
\ee
In terms of the two radial solutions $g_l(R)$ and $f_l(R)$, one can write \cite{Bini:2006dp} the electric scalar potential generated by a charge $q$ located at $z=D$ as
\be\label{vcomplete}
A_0=-V=-q\sum_l\left[f_l(D)g_l(R)\Theta(D-R)+f_l(R)g_l(D)\Theta(R-D)\right]P_l(\cos\theta),
\ee
where $\Theta(x)$ denotes the usual step function.\\
Inserting the result (\ref{vcomplete}) into the definition (\ref{sigmadef}) of the induced charge density (and using again the result (\ref{poldiff}) used above to compute the electric Love numbers) leads to the following result for the $l$-th multipolar component of the charge density induced on the surface of the BH
\be
4\pi R_0\sigma_l=-\f{l!(l+1)!}{(2l)!}qf_l(D)R_0^l=-h_l^{EM}qf_l(D)R_0^l.
\ee
This result can be interpreted in two ways. On the one hand, one can view it as an application of the general definition of Love numbers, Eq. (\ref{sigmal}), but with the understanding that, when the charge $q$ gets near the horizon, one must replace the usual ``EM tidal moment'' $\tau_l=-q/D^{l+1}$ by $\tau_l(D)=-qf_l(D)$. On the other hand, one can alternatively (in the spirit of Love's original definition) wish to continue to define $\tau_l$ as being simply the ``Coulombian'' values $\tau_l=-q/D^{l+1}$, in which case one can say that the Love number $h^{EM}$ in Eq. (\ref{sigmal}) must be \textit{dressed} by a distance-dependent correcting factor, say $t_l^{EM}(D)\equiv D^{l+1}f_l(D)$ (with $t_l(D)\rightarrow1$ as $D\rightarrow\infty$), so that $h_l^{EM}(D)=h_l^{EM}t_l^{EM}(D)$ plays the role of a distance-dependent effective Love number in
\be\label{hlemd}
4\pi R_0\sigma_l=-h_l^{EM}t_l^{EM}(D)\f{q}{D^{l+1}}R_0^l=-h_l^{EM}(D)\f{q}{D^{l+1}}R_0^l.
\ee  
Using Eq. (\ref{flr}) to evaluate $f_l(D)$ and remembering that \cite{whit} $Q_l(x)=\tfrac{1}{2}P_l(x)\log\f{x+1}{x-1}-\bar{P}_{l-1}(x)$, where $\bar{P}_{l-1}(x)$ is a polynomial of order $l-1$, one finds that the correcting factor $t_l^{EM}(D)\equiv D^{l+1}f_l(D)$ grows, as $D$ decreases from infinity down to $D=R_0=2M$, from $1$ to a horizon value equal to
\be\label{tl}
t_l^{EM}(2M)=\f{(2l+1)!}{l!(l+1)!}.
\ee
Inserting the result (\ref{tl}) into the definition of the ``dressed'', distance-dependent Love number leads to the following horizon value for the dressed Love number
\be\label{hl2M}
h_l^{EM}(2M)=h_l^{EM}t_l^{EM}(2M)=2l+1.
\ee
Note that, while the bare ``faraway'' Love numbers $h_l^{EM}$ \textit{decreased} roughly as $4^{-l}$ as $l$ increased, the dressed, near-horizon effective Love numbers (\ref{hl2M}) now \textit{increase} linearly in $l$. This linear increase with $l$ is the multipole-expanded version of the result found in Hanni and Ruffini \cite{Hanni:1973fn} that, as the charge $q$ (with, say, $q>0$) gets near the horizon, the induced charge density becomes nearly uniformly distributed on the horizon, apart from a strongly negative charge distribution, localized in a small patch ``below'' the charge $q$. Indeed, the limiting result (\ref{hl2M}) corresponds, when inserted in Eq. (\ref{hlemd}), to a charge distribution on the horizon given by (we recall that, by construction, the monopolar component $\sigma_0$ vanishes because the BH is uncharged)
\be\label{sigmaemnear}
\sigma=\sum_{l=1}^{\infty}\sigma_lP_l(\cos\theta)=-\f{q}{4\pi R_0^2}\sum_{l=1}^{\infty} (2l+1)P_l(\cos\theta)=-\f{q}{4\pi R_0^2}\left[2\delta(1-\mu)-1\right],
\ee
where $\delta(1-\mu)$ denotes a Dirac-delta distribution localized at $\mu=1^{-}$ (with the convention $\int_{a}^{1}d\mu\delta(1-\mu)=1, \forall a<1$). Here we have (formally) applied the general theorem on the Legendre expansion of a function $f(\mu)$ on the interval $\mu\in[-1,+1]$, namely 
\begin{subequations}
\begin{align}
&f(\mu)=\sum_{l=0}^{\infty}f_lP_l(\mu),\\
&f_l=\f{2l+1}{2}\int_{-1}^{+1}d\mu f(\mu)P_l(\mu)
\end{align}
\end{subequations}
to the distribution $f(\mu)=2\delta(1-\mu)$.\\
Note also that the result (\ref{hl2M}) coincides with the Love number of a conducting sphere in flat space, Eq. (\ref{conductingh}). This can be interpreted as meaning that, in the near-horizon limit, the electromagnetic interaction between the charge $q$ and the horizon becomes localized in a small patch below $q$, and that such a localized behavior applies also to the interaction between a charge in flat space which becomes very close to a conducting sphere.
\section{Gravitational case: near-horizon limit}
After this incursion into the electromagnetic analogs of tidal effects, let us come back to the gravitational case. We wish now to consider the gravitational analog of the Hanni-Ruffini study \cite{Hanni:1973fn}, i.e. the tidal deformation of the horizon by a test mass, $m\ll M$, in the limit where $m$ gets very close to the horizon. To do this, it is convenient to start from the general result derived above for the curvature of the horizon, namely
\be\label{4.1}
(2M_u)^2K(\mu)=-\f{1}{2}\partial^2_\mu\left[(1-\mu^2)e^{2(\bar{U}_{\rm ext}(\mu)-u)}\right].
\ee
Correlatively, the coefficients $c_l$ of the multipolar expansion of $K(\mu)$,
\be\label{4.2}
(2M_u)^2K(\mu)=1+\sum_{l=2}^{\infty}c_lP_l(\cos\theta),
\ee
(where we used our general results (\ref{c0}) and (\ref{c1}) above) are given by
\be\label{4.3}
c_l=-\f{2l+1}{4}\int d\mu P_l(\mu)\partial^2_\mu\left[(1-\mu^2)e^{2(\bar{U}_{\rm ext}(\mu)-u)}\right].
\ee
Here, $\bar{U}_{\rm ext}(\mu)$ is the value on the horizon of the Weyl-Newton external potential $U_{\rm ext}$, which is a solution of the axisymmetric, flat-space Laplace equation. We have in mind here a general configuration where $U_{\rm ext}$ is generated by a stationary axisymmetric configuration of masses, $m_i$, which include, among other sources, a test mass $m\ll M$ very close to the horizon.\\
Let us consider the limiting case where all the external masses are test masses: $m_i/M\rightarrow0$. In this limiting situation, $U_{\rm ext}$ ( and $u=U_{\rm ext}^{NP}=U_{\rm ext}^{SP}$) formally tends to zero, and we can replace Eqs. (\ref{4.1}) and (\ref{4.3}) by their linearized approximations:
\begin{subequations}
\begin{align}
&(2M_u)^2K^{\rm lin}-1=-\partial^2_\mu\left[(1-\mu^2)\left(\bar{U}_{\rm ext}(\mu)-u\right)\right], \qquad l\ge2,\\
&c_l^{\rm lin}=-\f{2l+1}{2}\int_{-1}^{+1} d\mu P_l(\mu)\partial^2_\mu\left[(1-\mu^2)\bar{U}_{\rm ext}(\mu)\right].\label{4.5}
\end{align}
\end{subequations}
In Eq. (\ref{4.5}) we have used the fact that, for $l\ge2$, the $\mu$-independent term $u=U^{NP}_{\rm ext}$ does not contribute to $c_l$. Finally, as $c_l^{\rm lin}$, Eq. (\ref{4.5}), is a \textit{linear} function of $U_{\rm ext}$, we can consider that each $c_l^{\rm lin}$ is given by a sum over the various external masses, say, $c_l^{\rm lin}=c_l^{\rm lin}(m)+\sum_ic_l^{\rm lin}(m_i)$. In the following, we shall focus on the contribution $c_l^{\rm lin}(m)$, where $m$ is a test mass which is very close to the horizon \footnote{Though the present linear approximation allows us to focus, independently from the other masses $m_i$, on the contribution $c_l^{\rm lin}(m)$, one should keep in mind that the various masses are not totally independent from each other as they must respect the ``equilibrium condition'' (\ref{NP=SP}), which is needed for ensuring the regularity of the metric on the portions of the $z$ axis  that touch the ``central'' BH of mass $M$.}. The individual contribution $c_l^{\rm lin}(m)$ is obtained by inserting in the right-hand side of Eq. (\ref{4.5}) the $\rho\rightarrow0$ limit of $-\psi_m$, where $\psi_m$ is given by Eq. (\ref{psim}). As we are considering the test-mass limit $m/M\rightarrow0$, we can simplify this logarithmic expression for $U_m=-\psi_m$ by expanding the logarithm in powers of $2m/(R_m^++R_m^-)$, and then by approximating $(R_m^++R_m^-)/2$ simply by $\sqrt{\rho^2+(z-b)^2}$. This yields the simple, Newton-like result
\be\label{4.6}
U_m\simeq\f{m}{\sqrt{\rho^2+(z-b)^2}}.
\ee
Then, taking the horizon limit (i.e. $\rho\rightarrow0$) and replacing $z$ by $M\mu$ according to Eq. (\ref{muesse}) leads to the explicit result (we assume, for definiteness, that $b>M>0$, i.e. that the mass $m$ is near the north pole, $\mu=+1$):
\be\label{4.7}
c_l^{\rm lin}(m)=-\f{2l+1}{2}\int_{-1}^{+1}d\mu P_l(\mu)\partial^2_\mu\left[\f{m(1-\mu^2)}{b-M\mu}\right].
\ee
There are now two ways to discuss what happens when $b$ decreases from $b\gg M$ down to $b\simeq M$. [We recall that $b$ is the Weyl-coordinate distance, so that the horizon is located at $b=M$.] A first way consists in noticing that the integral (\ref{4.7}) can be explicitly expressed in terms of the Legendre functions of the second kind. Indeed, using \cite{whit}
\be\label{4.8}
Q_l(x)=\f{1}{2}\int_{-1}^{+1}P_l(y)\f{dy}{x-y},
\ee
we can reexpress Eq. (\ref{4.7}) as
\be\label{4.9}
c_l^{\rm lin}(m)=(2l+1)\f{m}{M}(\hat{b}^2-1)\partial_{\hat{b}}^2Q_l(\hat{b}),
\ee
where $\hat{b}=b/M$.\\
Similarly to our discussion of the electromagnetic case, we can then re-interpret the result (\ref{4.9}) by writing the $l$-th multipolar component of the adimensionalized horizon curvature ``raised'' by an external test mass $m$ at Weyl distance $b$ in the form
\be\label{4.10}
\left[(2M_u)^2K^{\rm lin}\right]_l\simeq c_l^{\rm lin} (m)P_l(\mu)=(l-1)(l+2)h_lt_l(b)\f{m}{(b+M)^{l+1}}\f{R_0^{l+1}}{M}P_l(\mu),
\ee 
where $h_l$ is the bare, faraway gravitational Love number, Eq. (\ref{deltaR/R}), and where $t_l(b)$ is a distance-dependent correcting factor (normalized so that $t_l(b)\rightarrow1$ as $b\rightarrow+\infty$): here we conventionally replaced the factor $1/D^{l+1}$ in the electromagnetic definition (\ref{hlemd}) by $1/(b+M)^{l+1}$ to ensure that these ``Coulombian'' factors agree both when $b\rightarrow+\infty$ and on the horizon.\\
Comparing Eq. (\ref{4.9}) to Eq. (\ref{4.10}), we get 
\be\label{4.11}
(l-1)(l+2)h_lt_l(b)\left(\f{2M}{b+M}\right)^{l+1}=(2l+1)(\hat{b}^2-1)\partial_{\hat{b}}^2Q_l(\hat{b}).
\ee 
If we now formally\footnote{As our analysis here assumes $U_{\rm ext}\ll 1$, we should always keep $m/\epsilon\ll 1$. In other words, the coordinate distance to the horizon, $\epsilon=b-M$, should tend to zero less rapidly than $m/M$.
} let $\epsilon=b-M$ tend to zero, i.e. $\hat{b}=b/M\equiv1+\hat{\epsilon}$ (with $\hat{\epsilon}\equiv\epsilon/M$) tend to $1$, one finds that
\be\label{4.12}
(\hat{b}^2-1)\partial_{\hat{b}}^2Q_l(\hat{b})\simeq\f{1}{\hat{\epsilon}}\left[1-l(l+1)\hat{\epsilon}+\mathcal{O}(\hat{\epsilon}^2)\right],
\ee
the crucial point being that the coefficient of the leading term $\hat{\epsilon}^{-1}=(\hat{b}-1)^{-1}$ in the right-hand side is equal to $1$, independently from the value of $l$. In other words, the near-horizon limit of the multipolar coefficients of the curvature is ($\epsilon=b-M$)
\be\label{4.13}
c_l^{\rm lin}(m)|_{b\rightarrow M}=(2l+1)\f{m}{\epsilon},
\ee
and this result can be interpreted (similarly to the electromagnetic case (\ref{hl2M})) as coming from the horizon limit of a distance-dependent ``dressed'' Love number $h_l(b)$,
\be
h_l(M+\epsilon)=h_l\tau_l(M+\epsilon)\simeq\f{2l+1}{(l-1)(l+2)}\f{M}{\epsilon}.
\ee
In other words, the correcting factor $\tau_l(b)$ grows, as $b\rightarrow M$, in a strongly $l$-dependent manner (roughly $\sim4^{+l}$), and this growth compensates the usual behavior ($h_l\sim4^{-l}$) of the faraway Love number.
The second way to obtain the simple final result (\ref{4.13}) is to explicate the second $\mu$ derivative entering Eq. (\ref{4.7}) so as to obtain:
\begin{equation}
\begin{aligned}
&c_l^{\rm lin}(m)=(2l+1)\f{m}{M}\int_{-1}^{+1}d\mu P_l(\mu)\f{\hat{b}^2-1}{(\hat{b}-\mu)^3}\simeq\\
&\simeq(2l+1)\f{m}{\epsilon}\int_{-1}^{+1}d\mu P_l(\mu)\f{2\hat{\epsilon}^2}{(1+\hat{\epsilon}-\mu)^3}.\label{4.14}
\end{aligned}
\end{equation}
Then one notices that, in the near-horizon limit $\hat{\epsilon}=\epsilon/M\rightarrow0$, the function $\delta_{\hat{\epsilon}}(1-\mu)=2\hat{\epsilon}^2(1+\hat{\epsilon}-\mu)^{-3}=\partial_\mu[\hat{\epsilon}^2(1+\hat{\epsilon}-\mu)^{-2}]$ gets localized near $\mu=1^-$ and has an integral $\int_{a}^{+1}d\mu\delta_{\hat{\epsilon}}(1-\mu)=[\hat{\epsilon}^2(1+\hat{\epsilon}-\mu)^{-2}]_a^1=1-\hat{\epsilon}^2(1+\hat{\epsilon}-a)^{-2}$ which tends to $1$. In other words, $\delta_{\hat{\epsilon}}(1-\mu)\rightarrow\delta(1-\mu)$ as $\hat{\epsilon}\rightarrow 0$. Using $P_l(1)=1$, we then obtain another proof of the limiting result (\ref{4.13}).\\
Inserting the result (\ref{4.13}) in the multipolar expansion (\ref{4.2}), we then see that the contribution to $K$ coming from a mass $m$ tends, as $m$ approaches the horizon ($\epsilon=b-M\ll M$) towards
\be\label{4.15}
(2M_u)^2K^{\rm lin}(m)\simeq\f{m}{\epsilon}\sum_{l=2}^{\infty} (2l+1)P_l(\mu).
\ee  
Using, as we did in the near-horizon electromagnetic case above, the multipolar decomposition of the Dirac-delta distribution $2\delta(1-\mu)$, we can sum the series (\ref{4.15}) to get
\be\label{4.16}
(2M_u)^2K^{\rm lin}(m)\big|_{\epsilon\ll M}\simeq\f{m}{\epsilon}\left[2\delta(1-\mu)-1-3\mu\right].
\ee
This result is the gravitational analog of the electromagnetic result (\ref{sigmaemnear}). It shows that, modulo a global smooth curvature linear in $\mu$, the tidal horizon curvature generated by a near-horizon mass $m$ is localized in a small patch ``below'' the mass $m$. Similarly to the electromagnetic case, this result could also be formulated in saying that, in the near-horizon limit, the tidal curvature, being localized in a small patch on the horizon (as witnessed by the integrand $\propto(\hat{b}^2-1)/(\hat{b}-\mu)^3$ in Eq. (\ref{4.14})), could also be computed by replacing the real curved horizon by a flat-space horizon or ``Rindler horizon'' (as done in \cite{Suen:1988kq,Macdonald:1985}).\\
We will, however, get a more interesting result (which has no electromagnetic analog) by considering the ``shape distortion'' of the horizon associated to the tidal curvature (\ref{4.16}). Using our general result (\ref{cl}), the multipolar expansion of the corresponding near-horizon shape distortion is
\be\label{4.17}
\f{\delta R(\mu)}{R_0}=\f{m}{\epsilon}\sum_{l=2}^{\infty}\f{2l+1}{l(l+1)-2}P_l(\mu).
\ee 
To sum this new series, we apply the differential operator $\Delta+2$, where $\Delta$ denotes the Laplacian on the unit sphere (with $\Delta P_l(\mu)=-l(l+1)P_l(\mu)$) to Eq. (\ref{4.17}) and find that $\delta R(\mu)$ satisfies the differential equation
\be\label{4.18}
\f{\epsilon}{m}\left(\Delta+2\right)\f{\delta R}{R_0}=1+3\mu-2\delta(\mu-1),
\ee
where $\Delta$, acting on an axisymmetric function, can be replaced by $\partial_\mu(1-\mu^2)\partial_\mu$. This yields a second-order ordinary differential equation for $\delta R(\mu)$. The presence of a delta-function singularity on the right-hand side indicates that $\delta R(\mu)$ must have a $\sim\ln(1-\mu)$ singularity at $\mu\rightarrow1$ (and no singularity at $\mu\rightarrow-1$, contrarily to the $Q_0(\mu)$ Legendre function, which satisfies $(\Delta+2)Q_0=0$, modulo two delta-singularities at $\mu=1$ and $\mu=-1$). One then easily finds a particular solution of (\ref{4.18}) of the form  $\left(\Delta+2\right)\left(-\tfrac{1}{2}-\mu\ln(1-\mu)\right)=1+3\mu-2\delta(\mu-1)$. The general solution of (\ref{4.18}) is then obtained by adding to this particular solution a general regular solution of $(\Delta+2)f(\mu)=0$, i.e. a multiple of $P_1(\mu)=\mu$. One can determine this multiple by requiring (as needed from the absence of $l=0$ and $l=1$ contributions in (\ref{4.17})) that $\int d\mu P_1(\mu)\delta R(\mu)=0$. Finally, this uniquely determines the sum (\ref{4.17}) to be
\be\label{4.19}
\f{\delta R(\mu)}{R_0}=\f{m}{\epsilon}\left[-\f{1}{2}-\mu\ln(1-\mu)+\mu(\ln2-\f{4}{3})\right].
\ee
The corresponding shape is illustrated in Fig. 1 (for a prefactor $m/\epsilon=1/30$). This Figure shows the ``height'' of the tide raised on the horizon by a mass $m$, in the limit where the location $b$ of the mass tends to the horizon ($b\rightarrow M$, keeping $m/\epsilon=m/(b-M)$ finite and small). When $\epsilon$ is kept non zero (even beyond the prefactor $m/\epsilon$), one finds that the logarithmic ``spike'' at $\mu=\cos\theta=1$ (north pole) is rounded off on the characteristic angular scale $\theta_c\sim\sqrt{(b-M)/M}$, corresponding to the multipolar orders $l_c\sim\theta^{-1}_c\sim\sqrt{M/\epsilon}$. [The characteristic multipolar order $l_c$ can be read off the next-to-leading term in the expansion on the right-hand-side of Eq. (\ref{4.12}).]
\begin{figure}[ht]
\begin{center}
\includegraphics[width=0.5\textwidth]{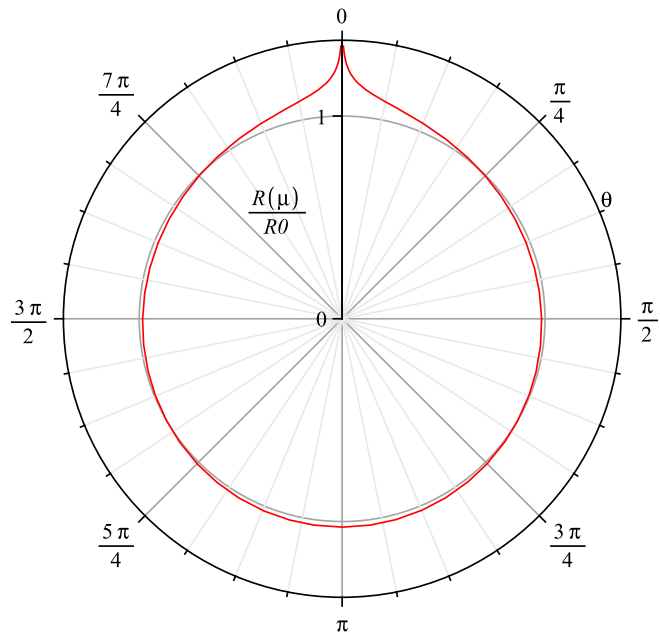}
\caption{The shape of a meridian section of the ``tidal bulge'' raised by a test-mass $m$ located at the north pole of the horizon of a BH.}
\end{center}
\end{figure}
\section{Conclusion}
We compared and contrasted the \textit{gravitational polarizability} properties of black holes (BH) with their \textit{electromagnetic polarizability} properties. Our main results are:
\begin{enumerate}
	\item The definition and computation of the infinite sequence of ``shape'' (or ``height'') Love numbers $h_l$ of a BH, where $l=2,3,4...$ is the multipolar order. The result is given by Eq. (\ref{hl}): $h_l$ essentially measures the ratio between the ``height''  of the $l$-th tidal bulge raised on the horizon of a BH and the corresponding external $l$-th tidal potential $U_{\rm ext}(R_0)$ (analytically continued down to the horizon $R_0=2GM$). Contrary to the Newtonian $h_l$ Love numbers of a perfect-fluid star which are larger than $1$ \cite{Damour:2009vw}, we found that the Love numbers of a BH are smaller than $1$, and tend rapidly (and exponentially) toward zero as $l$ increases, e.g. $h_2=1/4$, $h_3=1/20$, $h_4=1/84$,..., $h_l\simeq\sqrt{2\pi l}/2^{2l+1}$. In a related recent work \cite{Damour:2009vw}, it was found that the height Love numbers of a neutron star approach those of a BH as the ``compactness'' $c=GM/R$ of the star formally tends toward the BH value $c^{BH}=1/2$.
	\item The corresponding definition of a sequence of electromagnetic Love numbers $h_l^{EM}$ ($l=1,2,3,...$). These essentially measure the ratio between the $l$-th multipolar charge induced on the horizon of a BH and the corresponding external $l$-th multipolar electric potential (analytically continued down to the horizon). Here again we found that the electric Love numbers of a BH (given in Eq. (\ref{hlem})) are much smaller than those of a conducting sphere in flat space (\ref{conductingh}), and tend rapidly (and exponentially) toward zero as $l$ increases: $h_1^{EM}=1$, $h_2^{EM}=1/2$, $h_3^{EM}=1/5$, ..., $h_l^{EM}\simeq l\sqrt{2\pi l}/2^{2l}$. In addition, we found that they are simply related to the gravitational height Love numbers: $h_l^{EM}=2(l-1)h_l$.
	\item The comparative study of the gravitational and electromagnetic polarizability properties as the ``polarizing'' mass or charge approaches the horizon. Both cases can be described by replacing the ``bare'', faraway Love numbers, by some ``dressed'', distance-dependent factor $t_l(b)h_l$ or $t_l^{EM}(D)h_l^{EM}$. We found that the gravitational (respectively, electromagnetic) dressing factors $t_l(b)$ (respectively, $t_l^{EM}(D)$) compensate, in the near-horizon limit, the exponential decrease  of $h_l$ (resp. $h_l^{EM}$), and lead, in the gravitational case, to a specific ``spiky'' shape for the distorted  horizon  given by Eq. (\ref{4.19}), and illustrated in Fig. $1$.
\end{enumerate}
\section*{Acknowledgments} T.D. is grateful to David Ruelle for a discussion that prompted this investigation. We thank Alessandro Nagar for informative discussions. O.M.L. thanks IHES for its warm hospitality and gratefully acknowledges the support of both a Sapienza CUN 2 grant and an Angelo Della Riccia grant.

\end{document}